\documentclass[pre,showpacs,twocolumn,longbibliography]{revtex4-1}

\usepackage{hyperref}

\usepackage{color}
\usepackage[usenames,dvipsnames]{xcolor}
\usepackage{amsmath,amsthm,amssymb}
\usepackage{graphicx}
\usepackage{epsfig}
\usepackage{dcolumn}
\usepackage{bm}
\usepackage{mathrsfs}
\usepackage{multirow}
\usepackage[all]{xy}
\usepackage{pbox}
\usepackage{verbatim}
\usepackage[latin1]{inputenc}

\newcommand{\er}[1]{Eq.~\eqref{#1}}

\def\(({\left(}
\def\)){\right)}
\def\[[{\left[}
\def\]]{\right]}

\newcommand{\be}{\begin{equation}}
\newcommand{\ee}{\end{equation}}
\newcommand{\ben}{\begin{eqnarray}}
\newcommand{\een}{\end{eqnarray}}
\newcommand{\beq}{\begin{equation}}
\newcommand{\eeq}{\end{equation}}

\begin{document}

\title{Enhancing correlation times for edge spins through dissipation}

\author{Loredana M. Vasiloiu}
\author{Federico Carollo}
\author{Juan P. Garrahan}
\affiliation{School of Physics and Astronomy}
\affiliation{Centre for the Mathematics and Theoretical Physics of Quantum Non-Equilibrium Systems,
University of Nottingham, Nottingham, NG7 2RD, UK}

\date{\today}

\begin{abstract}
Spin chains with open boundaries, such as the transverse field Ising model,  
can display coherence times for edge spins that diverge with the system size 
as a consequence of almost conserved operators, the so-called strong zero modes. 
Here, we discuss the fate of these coherence times when the system is perturbed in two different ways. 
First, we consider the effects of a unitary coupling connecting the ends of the chain; when the coupling is weak and non-interacting, we observe stable long-lived harmonic oscillations between the strong zero modes.
Second, and more interestingly, we consider the case when dynamics becomes dissipative. While in general dissipation induces decoherence and loss of information, here we show that particularly simple environments can actually enhance correlation times beyond those of the purely unitary case. This allows us to generalise the notion of strong zero modes to irreversible Markovian time-evolutions, thus defining conditions for {\em dissipative strong zero maps}.
Our results show how dissipation could, in principle, play a useful role in protocols for storing information in quantum devices. 
\end{abstract}

\maketitle 

\section{Introduction}

Recent results have demonstrated the possibility of observing in many-body quantum chains with open boundary conditions  coherence times for edge spins that diverge exponentially with the size of the system \cite{Kemp,Fendley,Fendley1,Sarma,Alicea,Else}. 
One of the interests in this phenomenon stems from the possibility of storing and protecting the information encoded in quantum states for very long times, with possible applications in future quantum technologies 
\cite{Mazza,Bravyi}.
From this perspective, for the applicability of long coherence times for boundary degrees of freedom - so far only established in isolated (unitary) quantum systems \cite{Kemp,Fendley} - the effect of the inevitable interaction of the system of interest with the surrounding environment must be considered. This is the problem we address in this paper.

Acting as a source of dissipation and noise, the interaction of a system with an environment usually leads to suppression of truly quantum features in the system and to the emergence of classical-like behavior. However, in some instances it has been shown that engineered system-environment couplings can actually enhance or even generate quantum correlations, such as entanglement 
\cite{Benatti,Plenio,Braun,Kim,Schneider,Basharov,Jakobczyk,Benatti2,Benatti3,Kraus}.
While robustness of coherence times for edge spins has been extensively studied in the presence of interactions or of integrability-breaking terms and also in the presence of disorder in the Hamiltonian \cite{Kemp, Fendley,Lieu}, 
less is known about their behavior under irreversible open quantum dynamics \cite{Gardiner,Breuer,GARRAHAN,Lindblad,GKS}. In particular, it is not clear whether under Markovian (memory-less) dynamics these  long edge time-correlations can be observed and general conditions for their existence are not known. Understanding the behavior of these time-correlations in open quantum systems would allow for the possibility of  exploiting the protected information encoded in edge spins for applications in quantum devices  also in realistic non-equilibrium settings. 

\begin{figure}[t]
   \includegraphics[width=\columnwidth]{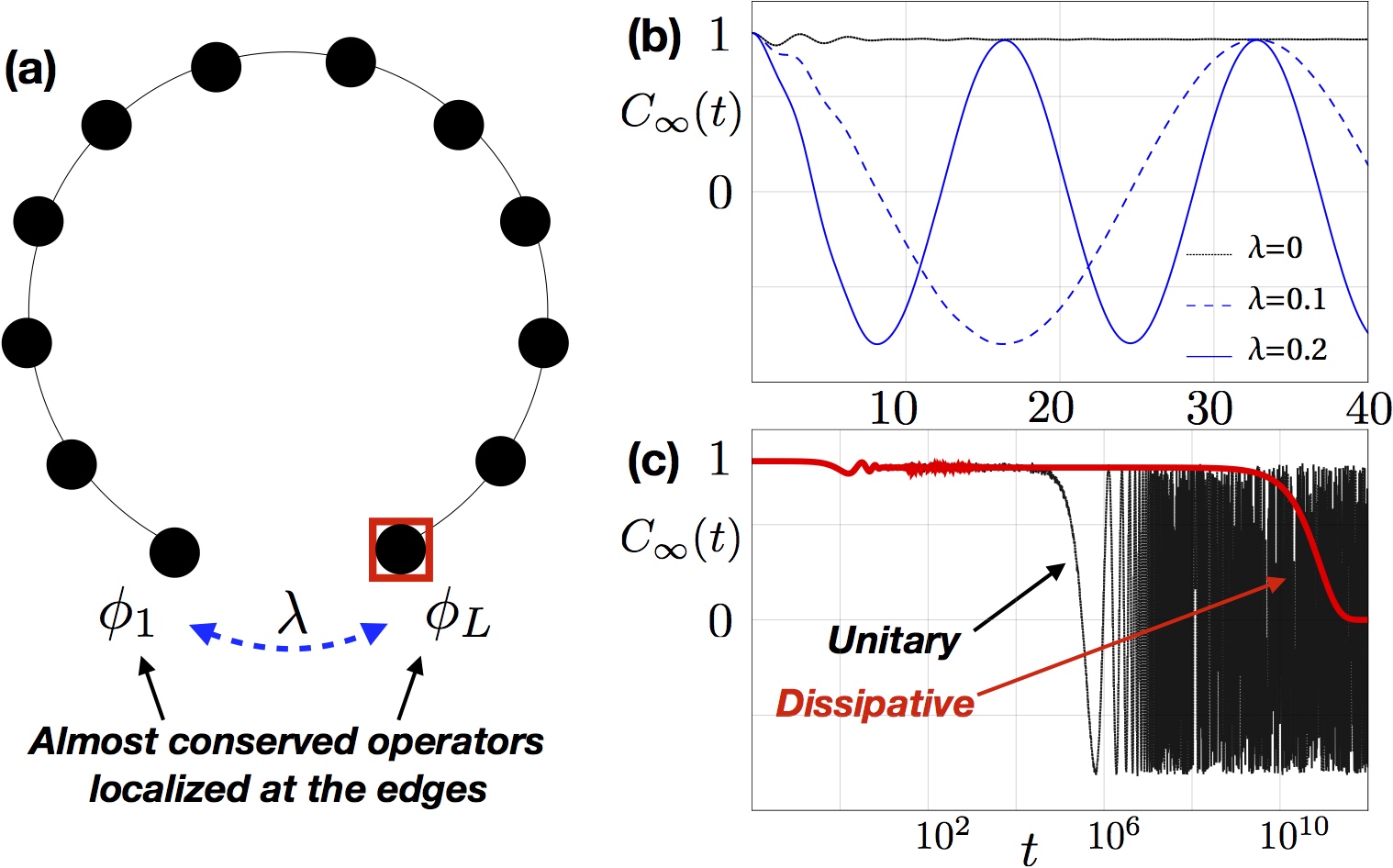}
  \caption{ {\bf (a)} Pictorial representation of a quantum spin chain featuring almost conserved operators, $\left( \phi_{1},\phi_{L} \right)$, localized at the two edges, perturbed either by weakly closing the chain (blue arrow) or by dephasing one edge (red box). 
  {\bf (b)} Behavior in time of the infinite temperature time-correlations of the $z$-magnetization of the first spin $C_\infty(t)$, given by Eq.~\eqref{AC},
  when the edges are joined through a non-interacting Hamiltonian perturbation $\left( \lambda \ne 0 \right)$. 
  Coherent oscillations between the two strong zero modes take place on faster time-scales than those of the free boundary case ($\lambda=0$). {\bf (c)}
  In the presence of dephasing  ($\Gamma \ne 0$)  acting only on one edge and $\lambda=0$ oscillations between the strong zero modes are suppressed; however the correlation time for the spin at the edge is doubled in order of magnitude. In the plot time is represented on a log-scale. }
   \label{Fig1}
\end{figure}

Long coherence times of edge spins are due to the presence of almost conserved operators -  the so-called {\it strong zero modes} (SZM)  \cite{Kemp,Fendley,Fendley1,Sarma,Alicea,Else}. The simplest example is that of the 
transverse field Ising model (TFIM) with open boundaries where the SZMs operators can be written down explicitly in compact form \cite{Kemp, Fendley1}. For finite sized systems, the SZMs are coupled by the dynamics but their evolution happens on very slow time-scales, so that initial information stored in the edge spins is preserved for times which are exponential in the system size \cite{Kemp}.

Here we consider the effect of perturbations away from the optimal conditions for the dynamical protection of the SZMs. One can think of two possible scenarios. The first one,  illustrated in Fig.~\ref{Fig1}(a), consists of connecting the two ends of an open chain via a unitary perturbation. Figure~\ref{Fig1}(b) illustrates our main result in this case:  
when the Hamiltonian perturbation is non-interacting (in a way we specify below) we observe  macroscopic coherent oscillations between the two SZMs, occurring on faster time-scales than in the disconnected case [cf.\ Fig.~\ref{Fig1}(c)] as the two ends are, in this case, directly connected. 

More interestingly, we then address the case in which the system undergoes a dissipative Markovian dynamics
induced by the presence of an external environment weakly interacting with it. 
Not only we show that Markovian quantum dynamics can sustain finite yet long correlation times, but also that suitably engineered - but still relatively simple - environments {\em can increase} coherence times by {\em orders of magnitude} with respect to the equilibrium closed system scenario. Figure~\ref{Fig1}(c) illustrates this result. In fact, these correlation times diverge with system size with a faster exponential rate in the dissipative case than in the purely coherent case. 

Indeed, as we will show, specific type of dissipations can induce fast decoherence effects affecting solely one of the two localized operators: 
coherent oscillations are suppressed but the information stored in one of the two edges is protected for much longer time-scales than the unitary ones. In particular, even a small perturbation, 
such as dephasing on the last site of the chain, can double in order of magnitude the correlation time for the first spin [see  Fig.~\ref{Fig1}(c)]. 
While for finite systems these correlations will eventually decay, in the thermodynamic limit they can persist for infinitely long times also in the presence of dissipation.

We mainly focus on the TFIM \cite{Sachdev,He} allowing for numerical diagonalization of large system size and amenable analytical considerations; 
nonetheless, we expect our findings to be very general and we show how they apply also for a dissipative interacting spin chain featuring almost conserved operators - the XYZ chain\cite{Baxter,Fendley}.

Finally, we introduce the conditions for the existence of  {\it strong zero maps}, which generalize the notion of  SZMs to irreversible Markovian quantum dynamics. 
These are obtained exploiting the possible presence of symmetry sectors in the dynamical generator of the open quantum dynamics \cite{Caspel,Buca}.\\

\section{Strong zero modes in the TFIM}

The simplest many-body system where the emergence of long correlation times for the edge spins can be observed is the TFIM  with free boundary conditions \cite{Sachdev,He}. It consists of a chain of two-level systems described by the Hamiltonian
\begin{equation}
H=-J \sum_{i=1}^{L-1} \sigma_i^{z} \sigma_{i+1}^{z}- h \sum_{i=1}^{L} \sigma_i^{x}\, ,
\label{HTFIM}
\end{equation}
where $\sigma^{\beta}_i$, with $\beta={x,y,z}$, represents the $\beta$ Pauli matrix corresponding to the $i$-th site. 
The parameter $J$ embodies the strength of magnetic interactions between neighboring sites, while $h$ is a transverse magnetic field in the $x$ direction. 
This Hamiltonian describes non-interacting fermions. Indeed, by introducing the Majorana operators
\begin{equation}
 \gamma_i^{A}= \sigma_{i}^{z} \prod_{k=1}^{i-1} \sigma_{k}^{x}, \; \; \; \; \gamma_i^{B}= \sigma_{i}^{y} \prod_{k=1}^{i-1} \sigma_{k}^{x}\, ,
\end{equation}
such that $\{\gamma^X_i,\gamma^Y_j\}=2 \delta_{X,Y}\delta_{i,j}$, the Hamiltonian can be written as
\begin{equation}
 H=-i J \sum_{i=1}^{L-1}  \gamma_i^{B} \gamma_{i+1}^{A}- i  h \sum_{i=1}^{L} \gamma_i^{A} \gamma_{i}^{B} .
\end{equation}
Introducing the parity operator
\begin{equation}
P=\prod_{i=1}^L\sigma^x_i ,
\end{equation}
it is immediate to check that $[H,P]=0$, implying that the eigenvectors of the Hamiltonian, as well as its eigenvalues, 
can be divided into even and odd sectors. Interestingly, in the magnetically ordered phase $\left| h \right| <\left| J \right|$, 
the even and odd parts of the Hamiltonian spectrum become identical exponentially fast in the system size $L$. 
This can be shown by introducing the so-called SZMs \cite{Kemp,Fendley,Fendley1,Sarma,Alicea,Else}
\begin{equation}
\phi_1= \mathcal{C} \sum_{i=1}^L \left( \frac{h}{J} \right)^{i-1} \gamma_{i}^{A} ,\; \; \; \; \phi_L= \mathcal{C} \sum_{i=1}^L \left( \frac{h}{J} \right)^{L-i} \gamma_{i}^{B} \label{SZMs} 
\end{equation}
where the constant ${\mathcal{C}}^2 =[1-\left(h/J\right)^2]/[1-\left(h/J\right)^{2L}]$ 
normalizes the operators in such a way that $\phi_{1/L}^2={\bf 1}$.
First of all, one notices that $\phi_1$ and $\phi_L$ anticommute with the parity operator $P$, and almost commute with the Hamiltonian $H$, 
\begin{equation}
[H,\phi_{1/L}]= \pm 2 i \mathcal{C} J \left( h/J\right)^L \gamma^{B/A}_{L/1} ,
\end{equation}
i.e., these commutators decay exponentially with system size. 
Therefore, with $|\psi_i\rangle$ being an even (odd) eigenvector of the Hamiltonian associated to the eigenvalue $\epsilon_i$, 
one has that $\phi_1|\psi_i\rangle$ is a vector which is odd (even) under parity transformation. Then, because $\|[H,\phi_1]\| \approx 0$ for large $L$, one has
$$
H\phi_1|\psi_i\rangle\approx\epsilon_i\;  \phi_1 |\psi_i\rangle \, ,
$$
showing that $\epsilon_i$ becomes, in the thermodynamic limit, an eigenvalue associated to both an even and an odd eigenvector. These two vectors are mapped one onto the other by the action of the SZM.

These properties of the Hamiltonian have remarkable practical consequences:
since the operators are localized at the two ends of the chain when $|h|\ll|J|$, the information in the boundary sites can be stored for times which are exponentially long in the system size \cite{Kemp}. 
Even more interestingly, this can actually be observed at high-temperatures as witnessed by the (infinite temperature) time-correlations 
\begin{equation}
C_\infty(t):=\frac{1}{2^L}{\rm Tr}\left({\rm e}^{-iHt}\sigma_1^z{\rm e}^{iHt}\sigma_1^z\right)\, .
\label{AC}
\end{equation}

\section{Weak bond in the ring geometry for the TFIM}

While the nice properties discussed above are completely lost when considering the same model in the ring geometry (i.e. the system is made translationally invariant through periodic boundary conditions), a question one may ask is what occurs
 in intermediate regimes that interpolate between the free boundary case with its long coherence times, and the ring geometry where they are absent. To this aim we introduce a Hamiltonian perturbation $H_{\rm b}$ to the TFIM Hamiltonian of Eq.~\eqref{HTFIM}, consisting of a term directly coupling the ends of the chain, as illustrated in Fig.\ref{Fig1}(a); the unitary time-evolution is then governed by $\hat{H}=H+\lambda H_{\rm b}$.

First, we consider a non-interacting perturbation $H_{\rm b}=-iJ\gamma_L^B\gamma_1^A$:
in this situation, for small $\lambda$, as we display in Fig.\ref{Fig1}(b), the time-correlations $C_\infty(t)$ manifest stable macroscopic oscillations. This feature emerges from the fact that the Hamiltonian perturbation connects harmonically 
the modes $\phi_{1/L}$, now with a finite (i.e. not decreasing with the system size) frequency that is isolated from the region of the spectrum that becomes dense in the large $L$ limit. 
Thus, while correlations of all other modes  are rapidly washed out by destructive interference, the two SZMs display stable  long-lived oscillations with frequency $\omega_{\lambda}=2 \lambda J $. When $\lambda\approx 1$, $\omega _\lambda$ enters in the continuous region of the spectrum and the stable oscillations fade away.
\begin{figure}
   \includegraphics[scale=0.35]{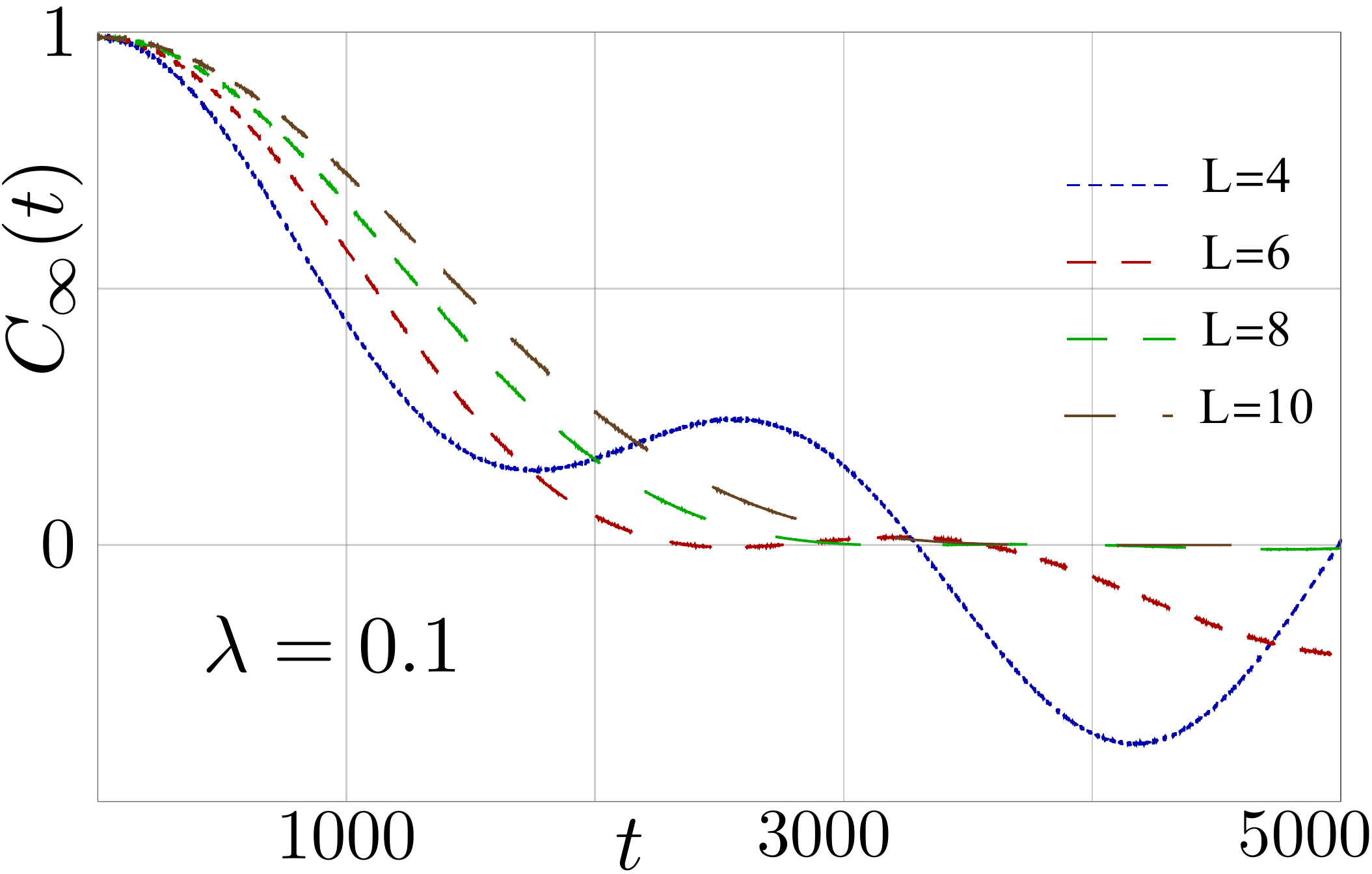}
  \caption{Behavior of the time correlation function $C_\infty(t)$ in the unitary case ($J=1$ and $h=0.1$) when joining the edges of the chain through the Hamiltonian $H_{\rm b}=-J\sigma_L^z\sigma_1^z$ with a small perturbative parameter $\lambda=0.1$. In this case, we see that $C_\infty(t)$ decays instead of undergoing macroscopic oscillations as shown in Fig.~\ref{Fig1}(b). For increasing system sizes the characteristic time of this decay increases, however it does not show an exponential dependence on $L$.  }
   \label{FigClosed}
\end{figure}
Completely different is the scenario in which boundary sites of the Ising chain are connected through the Hamiltonian $H_{\rm b}=-J\sigma_L^z\sigma_1^z$; 
in this case, the first-order perturbation correction vanishes and thus the modification to the evolution comes as a second-order map determining a decay of the correlation function $C_\infty(t)$ [see Fig.~\ref{FigClosed}].

\section{Dissipative dynamics in the open chain TFIM and enhancement of correlation times}

The results of the previous section concern instances of unitary quantum dynamics, thus describing a system which is perfectly isolated from its thermal surrounding. 
This situation is an idealized one and, in order to account for more realistic settings, one needs to consider open quantum evolutions \cite{Gardiner,Breuer,GARRAHAN,Lindblad,GKS}. For simplicity, we will consider the case of an environment weakly interacting with the system. Usually, such interaction leads to loss of quantum coherence and of quantum correlations.
However, as we now discuss, certain open quantum dynamics can still feature long correlation times for edge spins as a consequence of the existence of SZMs.

In the Markovian regime, the irreversible evolution of system observables is generated by $\dot{X}_t=\mathcal{L}^{*}[X_t]$ where $\mathcal{L}^*$ is the dual of the Lindbladian map $\mathcal{L}$ \cite{Lindblad,Gardiner,Breuer},  which generates the dynamics of the state of the system, $\rho$, through $\dot{\rho}=\mathcal{L}[\rho]$. The dual map has the following structure 
\begin{equation}
\mathcal{L}^{*}[X]:=i[H,X]+\sum_k\left(L_k^\dagger XL_k -\frac{1}{2}\left\{L_k^\dagger L_k, X\right\}\right)\, .
\label{LG}
\end{equation}
In the generator above, the first term represents the coherent part of the dynamics while the terms in the sum encode noisy effects due to the environment. 
We now assume that the presence of the environment induces decoherence in the spin chain, which can be  described by Lindblad operators 
\begin{equation}
L_i=\sqrt{\Gamma_{i}}\sigma_i^z, 
\label{Li}
\end{equation}
with $\Gamma_{i}$ being the dephasing rate at site $i$.

\begin{figure}
   \includegraphics[scale=0.34]{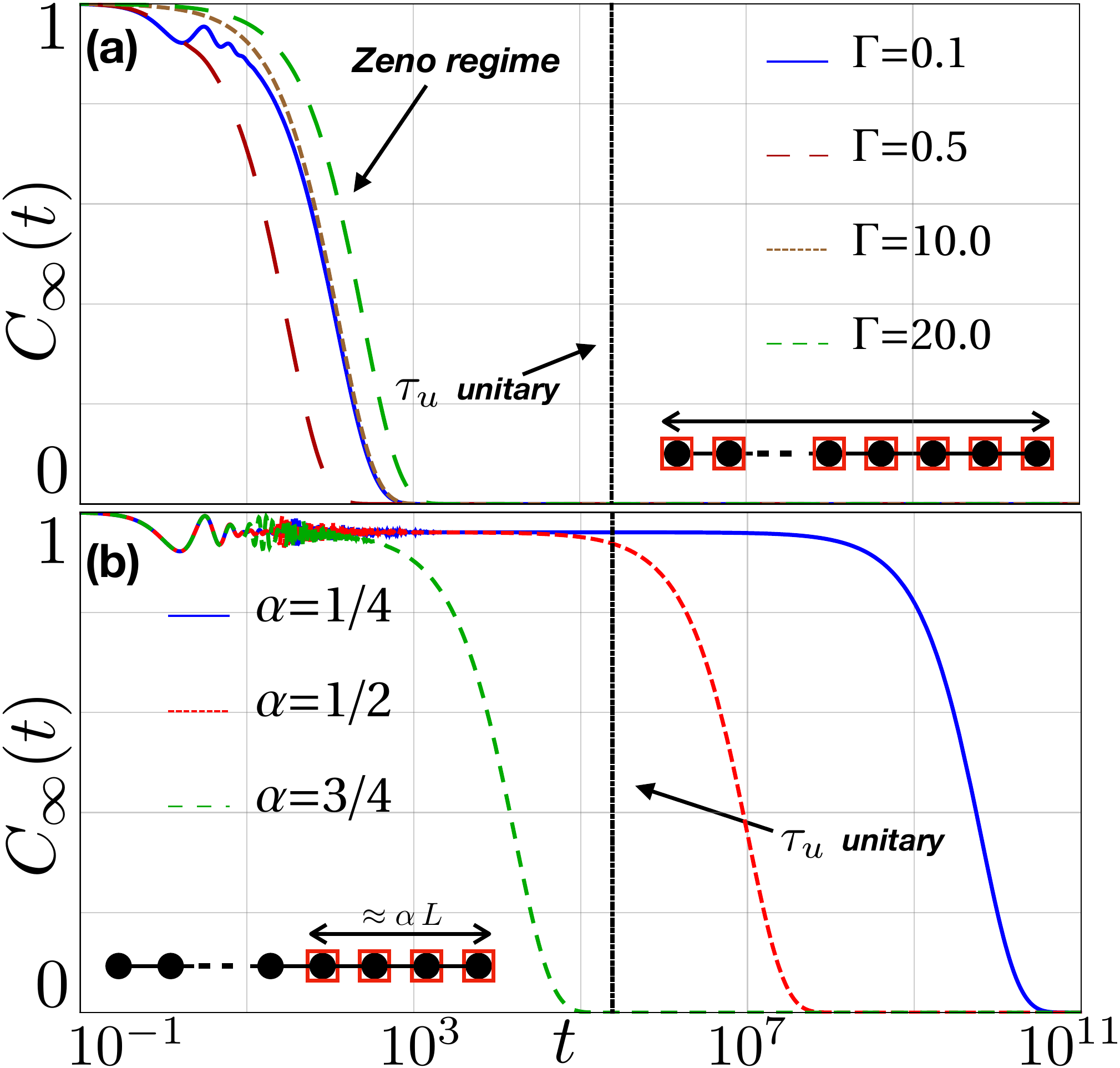}
  \caption{ Transverse field Ising model subject to dephasing on a fraction $\alpha$  of the chain starting from the last site. 
  {\bf (a) } For $\alpha=1$ the infinite temperature time-correlation $C_{\infty} \left( t \right)$ decays much faster than in the unitary case (the plot is for $J=1$, $h=0.2, L=8$ and different $\Gamma$). When $\Gamma\gg1$ the time-scales of the decay of $C_\infty(t)$ are dominated by the Zeno effect and the presence or not of a strong zero mode is irrelevant. 
  {\bf (b) } In the presence of dissipation only on a small portion of the chain, $\alpha < 1$, 
  we observe that $C_\infty(t)$ can decay later than in the unitary case (data shown for the same parameters $J$, $h$ and $L$ as in  {\bf (a)} with $\Gamma=1$ and different $\alpha$.) }
   \label{Fig2}
\end{figure}

As one would expect, when dephasing acts uniformly on the whole system, $\Gamma_{i} = \Gamma$, the spin-chain cannot sustain correlation times that grow with system size [see Fig.\ref{Fig2}(a)]. Nevertheless, with $\Gamma$ small, the presence of SZMs of the coherent case, 
leads to correlations times for the boundary spins which, while finite, are longer than those where SZMs are not present (for instance when $h=J$). 
When $\Gamma \gg1$ the physics changes, and time-correlations increase irrespective of the presence or the absence of SZMs in the unitary case. This corresponds to a quantum Zeno regime, with characteristic decay time given by $\tau_{\rm z}\approx \Gamma/(2h^2)$, which can be obtained straightforwardly from second-order perturbation theory.

A very different scenario emerges when one considers dissipation only on a fraction $\alpha$ of the chain, 
\begin{equation} 
\left\{
\begin{array}{cc}
\Gamma_{i} = 0 & i < L (1-\alpha) \\
\Gamma_{i} = \Gamma & i \geq L (1-\alpha) \\
\end{array}
\right.
\label{Ga}
\end{equation}
as illustrated in Fig.\ref{Fig2}(b). 
Also in this dissipative case we see that characteristic decay times for $C_\infty(t)$  are exponentially large with the size of the system. 
This is a surprising result suggesting that SZMs and infinitely long time-correlations can be observed in open quantum dynamics. 

However, there is a much more interesting effect that we can observe: not only, as we already pointed out, SZMs can exist in dissipative settings, 
but, strikingly, the characteristic decay time of  time-correlations associated to their existence can be enhanced by the presence of an external environment. 
This is clearly displayed in Fig.~\ref{Fig2}(b): decreasing the portion of the chain affected by dephasing - which we denote by $\alpha\in[0,1]$, cf.\ \eqref{Ga} - we can see that the time-correlations $C_\infty(t)$ stay almost invariant for larger times and, interestingly, for certain values of $\alpha$ these can be much greater than the unitary characteristic time $\tau_u$, where $\tau_u$ is the first instance when $C_\infty(t)=1/{\rm e}$. 

\begin{figure}
   \includegraphics[scale=0.4]{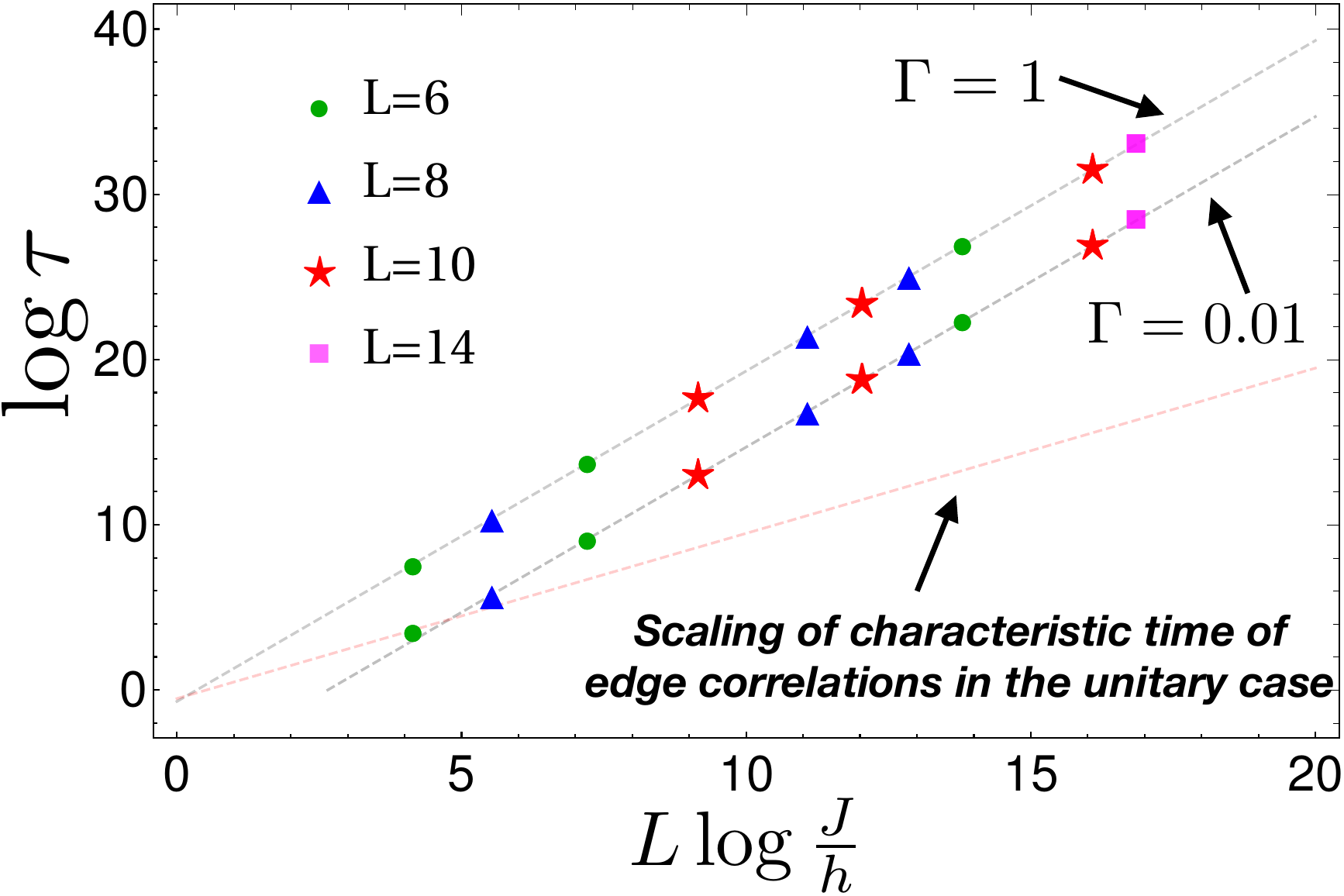}
  \caption{Behavior of the logarithm of the characteristic time of the correlation functions $C_\infty(t)$ as a function of $x=L\log (J/h)$. In the unitary case the scaling is characterized by slope equal to one ($\log \tau \approx x$). Remarkably, the presence of dephasing affecting the last site of the chain, enhances the scaling of $\log\tau$ which clearly shows, in this case, a behavior $\log \tau \approx 2x$, meaning that correlation time is doubled in order of magnitude with respect to the unitary case. In the plot we fix $J=1$ and explore different values of $L$, $h$ and $\Gamma$.}
   \label{Fig3}
\end{figure}

In order to understand this feature we consider the extreme case in which only the last site of the chain is subject to dephasing. Notice that this represents, in the large L limit, a perturbation to the unitary time-evolution which is infinitely far apart from the SZM $\phi_1$ which is instead localized around the first edge. In this scenario, considering a small transverse field $h$, the effective ``slow'' dynamics of the SZMs is given, at the largest order in $h$,  by the following system of differential equations
\begin{equation}
\frac{d}{dt}\begin{pmatrix}\phi_1 \\ \phi_L\end{pmatrix}=\begin{pmatrix}0&-2J\left(\frac{h}{J}\right)^L\\ 2J\left(\frac{h}{J}\right)^L &-2\Gamma\end{pmatrix}\begin{pmatrix}\phi_1 \\ \phi_L\end{pmatrix}\, .
\label{effdyn}
\end{equation}
Computing the eigenvalues of the matrix in the above equation one finds $\Delta_\pm=-\Gamma \pm \sqrt{\Gamma^2- 4 J^2 \left( h/ J \right)^{2L}}$; for increasing $L$ one always reaches a regime where $\Gamma\gg (h/J)^L$ and thus we can expand the square root in order to obtain a prediction for the characteristic decay time 
\begin{equation}
\log \tau\approx \log\frac{\Gamma}{2J^2}+2L\log \frac{J}{h}\, .
\label{taustima}
\end{equation}
Noticing that for the unitary case one expects $\log \tau_u\propto L\log(J/h)$\cite{Kemp}, we immediately see how Eq.\eqref{taustima} predicts, for the dissipative case under investigation, a characteristic time for the edge correlation function $C_\infty(t)$ which is doubled in order of magnitude with respect to the unitary evolution. In Fig.~\ref{Fig3} we show how the prediction in Eq.\eqref{taustima} is confirmed by numerical results.

While it is remarkable that a small modification of the dynamics affecting a site far apart from the first edge can have such a strong effect on $C_\infty(t)$, the same scaling of the correlation times can also be found for a dissipative dynamics affecting the whole chain and described by jump operators given by $\sigma_k^{z}\sigma^y_{k+1}$ for $k=1,2,\dots L-1$.

\section{Dissipative strong zero maps}

In this section we now proceed to a formal definition of SZMs in dissipative contexts.
This is achieved by promoting the notion of SZM from an operator acting on vectors to that of a map acting on operators. Let us consider a Lindblad generator $\mathcal{L}$ as in equation \eqref{LG} and assume that it commutes with a map implementing a discrete symmetry transformation on the system operators. Given the generator $S$ of the symmetry, the map can be written as $\pi_s[X]=SXS$; thus if the operator $X$ is even under this transformation one has $\pi_s[X]=X$, while if it is odd one has $\pi_s[X]=-X$.  The fact that the Lindblad operator commutes with the transformation $\mathcal{L}^{*}\circ \pi_s=\pi_s \circ\mathcal{L}^{*}$ means that one can divide eigenmatrices and eigenvalues of $\mathcal{L}$ into even and odd sectors. If there exists a map $\Psi$ commuting, up to exponentially small corrections in system size, with the Lindblad generator  
\begin{equation}
\mathcal{L}^{*}\circ\Psi[X]\approx \Psi\circ \mathcal{L}^{*}[X]\, ,\qquad \forall X\, ,
\end{equation}
and anti-commuting with the parity transformation 
\begin{equation}
\Psi\circ \pi_s=-\pi_s\circ\Psi\, ,
\end{equation}
then we can show that the even part of the spectrum of the Lindblad operator $\mathcal{L}$ is exponentially (in system size) close to the odd part. Indeed, considering $X_+$ to be an even eigenmatrix of $\mathcal{L}$, namely
\begin{equation}
\pi_s(X_+)=X_+\, ,\qquad \mathcal{L}^{*}[X_+]=\epsilon_+X_+\, ,
\end{equation}
one has that $\Psi[X_+]$ is an odd operator under the transformation, $\pi_s\circ\Psi[X_+]=-\Psi[X_+]$, and, because of the almost commutation of $\Psi$ with $\mathcal{L}$, also
$$
\mathcal{L}^{*}\Big[\Psi[X_+]\Big]\approx \Psi\Big[\mathcal{L}^{*}[X_+]\Big]=\epsilon_+\Psi[X_+]\, .
$$
\begin{figure}[t]
   \includegraphics[scale=0.22]{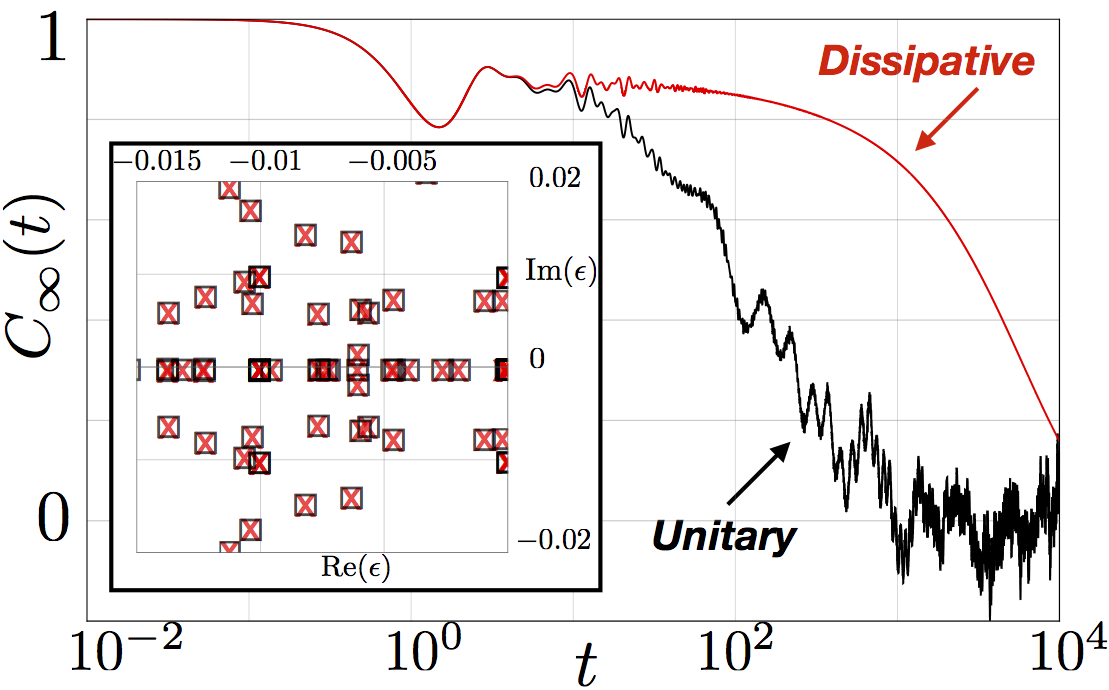}
  \caption{ Time correlation function $C_\infty(t)$ for the XYZ chain ($L=8$) with $J_x=0.2,\, J_y=0.3,\, J_z=1$ in the unitary case as well as with dephasing $\Gamma=1$ on the right boundary. 
  Also for this interacting Hamiltonian we see that the presence of dephasing enhances the characteristic correlation time for the first spin.
  In the inset, we display a portion of the spectrum for the XYZ chain ($L=6$) with $J_x=J_y=0.1$ and $J_z=1$: crosses represent eigenvalues from the odd sector while squares are from the even one. We found the largest distance between corresponding eigenvalues in the two sectors to be of order $10^{-4}$.   }
   \label{Fig4}
\end{figure}

This shows that $\Psi[X_+]$ is, in the thermodynamic limit, an odd eigenmatrix of $\mathcal{L}$ with eigenvalue $\epsilon_+$, meaning that the odd and even parts of the spectrum of the Lindblad operator are indeed paired. In order for the above derivation to be meaningful it is important that $\|\Psi[X]\|=\|X\|$, $\forall X$ or, at least, that the norm $\|\Psi[X]\|$ is neither zero nor divergent for all operators $X$. As a consequence of this pairing, with dissipative quantum dynamics (and not just coherent dynamics), 
correlation times of system observables which diverge exponentially with system size 
can be observed at infinite temperature. 

For example, considering the operator $\Psi[{\bf 1}]$ where ${\bf 1}$ is the identity, one has 
\begin{align}
\frac{1}{2^L}{\rm Tr}\left({\rm e}^{t\mathcal{L}^{*}}\Big[\big(\Psi[{\bf 1}]\big)^\dagger\Big]\Psi[{\bf 1}]\right) &\approx \frac{1}{2^L}{\rm Tr}\left(\Big(\Psi\Big[{\rm e}^{t\mathcal{L}^{*}}[{\bf 1}]\Big]\Big)^\dagger\Psi[{\bf 1}]\right) \nonumber \\
&=\frac{1}{2^L}{\rm Tr}\left(\Big(\Psi[{\bf 1}]\Big)^\dagger \Psi[{\bf 1}]\right)\, .
\label{DisAcorr}
\end{align}
In the examples of the previous section, the strong zero map is given by $\Psi[X]=\phi_{1/L}X$ or equivalently by $\Psi[X]=X\phi_{1/L}$, and $S=P$.\\

\section{Strong Zero Map in an interacting system}

The construction of the dissipative strong zero maps (DSZMs) of the previous section is generic and does not depend on a specific model. So far, however, we have only discussed the case of the TFIM which is in effect a system of non-interacting fermions. To show the more general applicability, we now study the case of an XYZ chain with open boundaries which, in contrast to the TFIM, corresponds to an interacting system. 

The Hamiltonian we consider is
\begin{equation}
H_{XYZ}=\sum_{k=1}^{L-1}\sum_{\beta=x,y,z}J_\beta\sigma_k^\beta\sigma_{k+1}^\beta \, ,
\label{Hxyz}
\end{equation}
with couplings $J_{\beta}$ which in principle can be different for the three components $\beta=x,y,z$. 
Furthermore, we consider dissipative dynamics, \er{LG}, with dephasing acting on the last site of the lattice, that is, with jump operators \eqref{Li} with $\Gamma_{i} = \Gamma \delta_{i,L}$. 

Figure~\ref{Fig4} shows the comparison between the boundary correlation function $C_\infty(t)$ for the case where dynamics is unitary generated by \er{Hxyz} and the case where there is dephasing on the last site. From Ref.~\cite{Fendley} we know that  in the coherent case, the existence of a SZM in the XYZ model gives rise to long coherence time (see black curve in Fig.~\ref{Fig4}), for analogous reasons to what occurs in the TFIM case. The corresponding correlator in the dissipative case is shown as the red curve in Fig.~\ref{Fig4}. Due to the existence of a DSZM, as in the TFIM case (cf.\ Fig.~\ref{Fig2}), correlation times are significantly enhanced due to dissipation at the opposite edge. The inset to Fig.~\ref{Fig4} show that indeed the (complex) spectrum of the Lindbladian is paired (approximately for the finite system shown, but the splitting will vanish exponentially in system size) between even and odd eigenstates of 
$\mathcal{L}$.

\section{Conclusions} 

We have considered the stability of strong zero modes in quantum chains with open boundaries. Our central result is that local dissipation can significantly enhance the correlation time of boundary spins. On the one hand this is surprising, as dissipation often acts to suppress memory of initial states as it tends to open up relaxation channels beyond those existing in its absence. We find that for both the non-interacting transverse field Ising model and for the interacting XYZ model, dephasing at one end of an open chain increases the correlation time of spins at the opposite end. This observation led us to define disspative strong zero maps, corresponding to superoperators that play the role in the case of dissipative (and Markovian) dynamics of the strong zero mode operators present in the unitary case. It would be interesting to probe our results by means of chains of Rydberg atom or trapped ion systems \cite{Bohnet,KimChang,Schaub,Bernien,Labuhn}.

\acknowledgments

The research leading to these results has received funding from a VC Scholarship for Research Excellence from the University of Nottingham (LMV), the European Research Council under the European Union's Seventh Framework Programme (FP/2007-2013) / ERC Grant Agreement No. 335266 (ESCQUMA) (FC), and from the EPSRC Grant No. EP/M014266/1 (JPG).

\bibliography{SZM}
\bibliographystyle{apsrev4-1} 
\end{document}